\documentclass[aps,prb,superscriptaddress,twocolumn,10pt]{revtex4-1}
\usepackage{amsfonts}
\usepackage{amsmath}
\usepackage{amssymb}
\usepackage[caption=false]{subfig}
\usepackage{caption}
\usepackage{tabto}
\usepackage{graphicx}
\usepackage{physics}
\usepackage{siunitx}
\usepackage{alltt}
\usepackage{textgreek}
\usepackage{textcomp}
\usepackage{nicefrac}
\usepackage{gensymb}
\usepackage{fancyhdr}
\usepackage[ddmmyyyy]{datetime}
\usepackage[printonlyused]{acronym}
\usepackage{color, colortbl}
\usepackage{textcomp}
\usepackage{array,multirow}
\usepackage{comment}
\usepackage{float}
\usepackage{booktabs}
\usepackage{bm}
\usepackage{array}
\usepackage{tabu}
\usepackage{gensymb}
\usepackage{hhline}
\usepackage[normalem]{ulem}
\newcommand{\up}[1]{\textsuperscript{#1}}

\definecolor{Gray}{gray}{0.8}
\definecolor{Grayer}{gray}{0.85}
\definecolor{Blue}{rgb}{0.8,0.8,1.0}
\definecolor{Bluer}{rgb}{0.6,0.6,1.0}
\definecolor{Yellow}{rgb}{1.0,1.0,0.8}
\definecolor{Oranzj}{rgb}{1.0,0.5,0.0}
\definecolor{Oranzj2}{rgb}{0.4,0.2,0.0}
\definecolor{Pinkz}{rgb}{1.0,0.5,0.75}
\definecolor{Redz}{rgb}{1.0,0.0,0.0}
\definecolor{Purple}{rgb}{0.65,0.0,0.65}

\newcommand{\grsq}{$\textcolor{green}{\blacksquare}$\ }
\newcommand{\yesq}{$\textcolor{yellow}{\blacksquare}$\ }
\newcommand{\orsq}{$\textcolor{Oranzj}{\blacksquare}$\ }
\newcommand{\resq}{$\textcolor{red}{\blacksquare}$\ }
\newcommand{\pusq}{$\textcolor{Purple}{\blacksquare}$\ }
\newcommand{\blsq}{$\textcolor{black}{\blacksquare}$\ }
\definecolor{Green}{rgb}{0.8,1.0,0.8}

\begin{document}
\title{Towards a simplified description of thermoelectric materials:
  Accuracy of approximate density functional theory for phonon dispersions}
\author{Thomas A. Niehaus}
\affiliation{ Univ Lyon, Universit\'e Claude Bernard Lyon 1, CNRS,
   Institut Lumi\`ere Mati\`ere, F-69622, Villeurbanne, France.}
\email{thomas.niehaus@univ-lyon1.fr}
\author{Sigismund T.A.G. Melissen}
\affiliation{ Univ Lyon, Universit\'e Claude Bernard Lyon 1, CNRS,
   Institut Lumi\`ere Mati\`ere, F-69622, Villeurbanne, France.}
\author{Balint Aradi}
\affiliation{BCCMS, University of Bremen, 28359 Bremen, Germany}
\author{S. Mehdi Vaez Allaei}
\affiliation{Department of Physics, University of Tehran,Tehran, Iran}

\begin{abstract}
We calculate the phonon-dispersion relations of several
two-dimensional materials and diamond using the density-functional based tight-binding approach
(DFTB). Our goal is to verify if this numerically efficient method
provides sufficiently accurate phonon frequencies and group velocities
to compute reliable thermoelectric properties.  To this end, the
results are compared to available DFT results and experimental
data. To quantify the accuracy for a given band, a descriptor is introduced that
summarizes contributions to the lattice conductivity that are
available already in the harmonic approximation. We find that the DFTB
predictions depend strongly on the employed repulsive pair-potentials,
which are an important prerequisite of this method. For carbon-based
materials, accurate pair-potentials are identified and lead to
errors of the descriptor that are of the same order as differences
between different local and semi-local DFT approaches.         
\end{abstract}
\date{\today}
 \maketitle

\begin{figure*}
\subfloat[\textit{h}-BN]{%
  \includegraphics[clip,width=0.35\columnwidth]{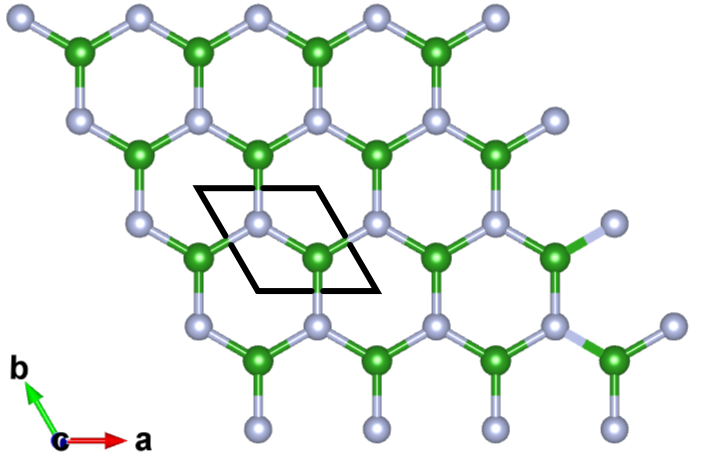}%
}
\hfill
\subfloat[Graphene]{%
  \includegraphics[clip,width=0.35\columnwidth]{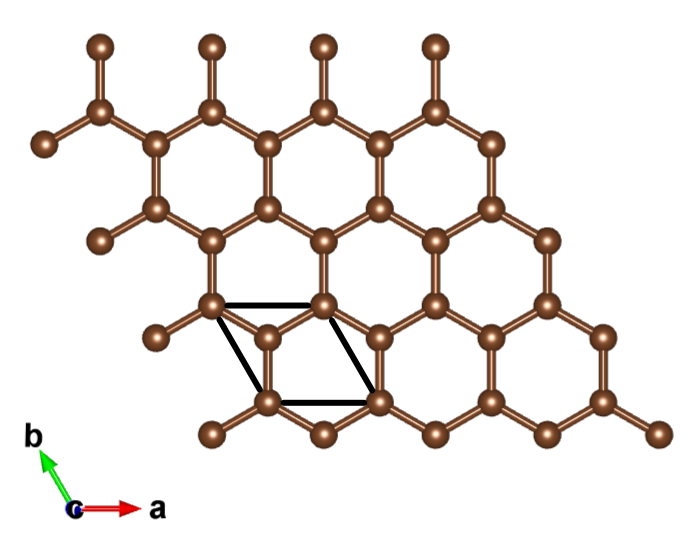}%
}
\hfill
\subfloat[Graphane]{%
  \includegraphics[clip,width=0.35\columnwidth]{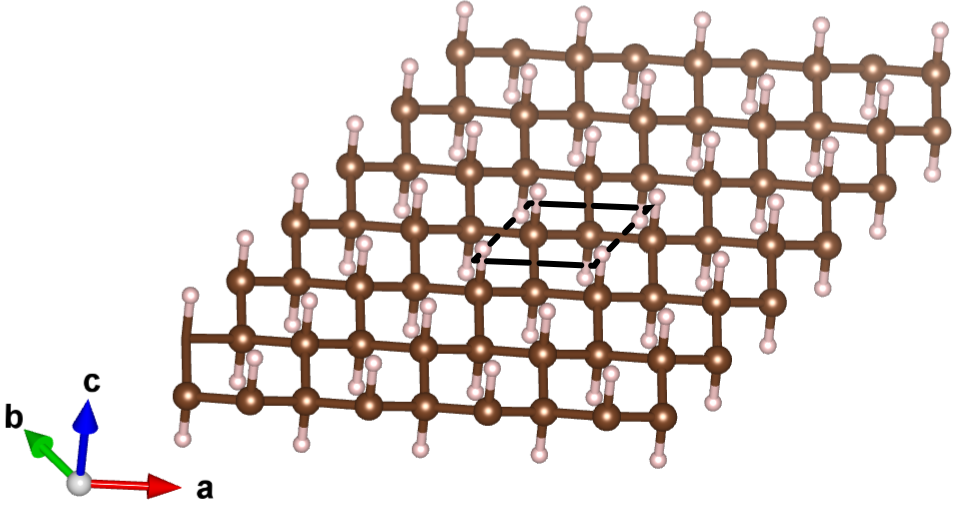}%
}

\subfloat[Diamond]{%
  \includegraphics[clip,width=0.35\columnwidth]{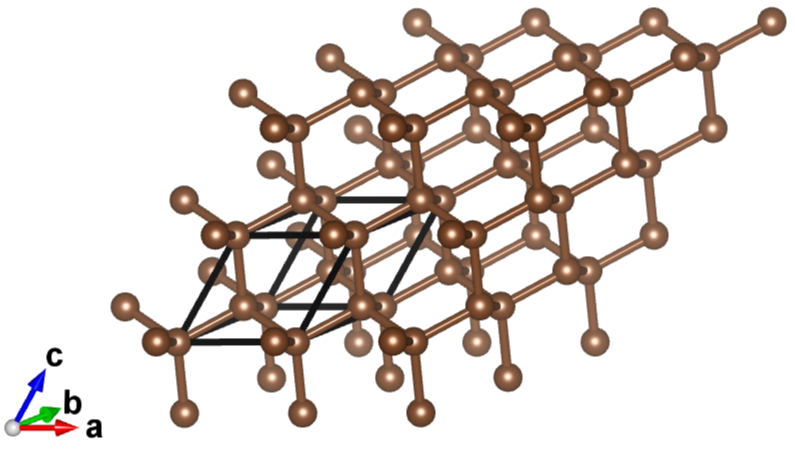}%
}
\hfill
\subfloat[Blue phosphorene]{%
  \includegraphics[clip,width=0.35\columnwidth]{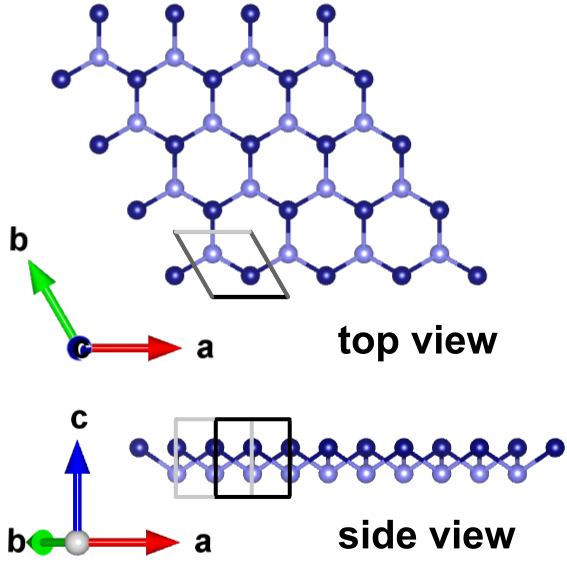}%
}
\hfill
\subfloat[Black phosphorene]{%
  \includegraphics[clip,width=0.35\columnwidth]{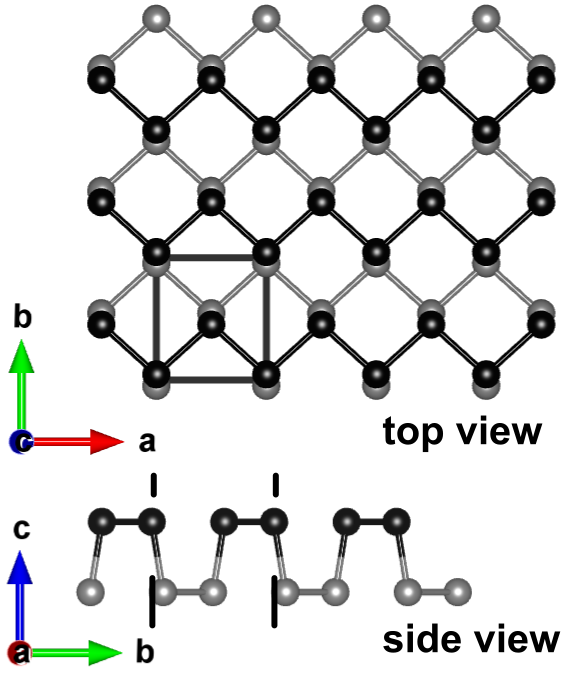}%
}
 \caption{Crystal structures of the compounds studied in this work. \label{struct}}
\end{figure*}

\section{Introduction}

The direct conversion of a temperature gradient to electric voltage or vice versa is known as the thermoelectric effect.
Although first rigorously defined following a series of discoveries in the mid 19\up{th} century, it was not until the mid 20\up{th} century that materials exhibiting interesting thermoelectric properties were sufficiently understood to enable targeted research.
Today, anthropogenic waste heat contributes significantly to climate change and for economic, as well as environmental reasons, creates a strong imperative to develop new thermoelectric materials.\cite{TanChemRev2016}
One feature of new materials often being exploited for potential thermoelectric applications is their anisotropy.
Indeed, among the currently best performing materials are layered materials.\cite{Snyder_Nat_Mater_2008}
The low hanging fruits of science inevitably being picked first, such materials become ever more complex and contain an ever greater variety of chemical elements.
The prediction of such materials' fundamental thermoelectric
properties using theoretical calculations prior to their synthesis can help experimentalists make specific choices in their target materials and, when such materials behave differently than predicted, such calculations can serve as a diagnostic tool.\cite{YangAdvEnMater2013}\\
The characteristic figure of merit for thermoelectric materials is
typically denoted $ZT$, and defined as:
\begin{equation}
ZT=\frac{\sigma S^2 T}{\kappa}
\label{ZT}
\end{equation}
with $T$ the temperature (K), $\sigma$ the electrical resistivity (S
m\up{-1}), $S$ the ``thermopower'' or Seebeck coefficient (V K\up{-1})
and $\kappa$ the thermal conductivity (W m\up{-1} K\up{-1}). In
crystalline materials the thermal conductivity is typically divided
into 
contributions from electrons ($\kappa_E$) and phonons ($\kappa_L$),
such that the total conductivity is the sum of both: $\kappa =
\kappa_E + \kappa_L$.

According to the Boltzmann transport equation,
the phonon (or lattice) conductivity along a certain crystallographic
direction $\alpha$ 
can be further broken down to\cite{Li2014}
\begin{equation}
  \label{klat}
\kappa_L^\alpha = a(T) \sum_j \sum_{\bf q}
f (f+1) v_{j\alpha}^2({\bf q}) [\hbar \omega_{j}({\bf q})]^2
\tau_{j}({\bf q}),   
\end{equation}
where $a(T)$ is a temperature dependent prefactor, $f$ 
denotes Bose-Einstein occupation factors, and $\omega_{j}({\bf q})$ and
$\tau_{j}({\bf q})$ correspond respectively to the frequency and lifetime
of a phonon in band $j$ at wave-vector ${\bf q}$. Finally, $v_{j\alpha}= \partial
\omega_{j} / \partial q_\alpha$ stands for the phonon group
velocity. While one needs to go beyond the harmonic approximation to obtain
the phonon lifetimes\cite{Togo2015}, both $\omega_{j}$ and
$v_{j\alpha}$ are readily available from the phonon band
structure (BS). An accurate description of the BS is therefore key to a
reliable computation of the lattice conductivity, and by means of
Eq.~\ref{ZT}, also ZT.

First-principles determinations of the phonon BS are typically based
on Density Functional Theory (DFT) which is the method of choice for
systems with unit cells comprising several tens of atoms. Given that
$ZT$ is inversely proportional to $\kappa$, recent attempts to increase
thermoelectric effiency make use of nanostructured materials that
feature extended structural defects \cite{Yamawaki2018} or complex unit cells
\cite{Snyder_Nat_Mater_2008} to suppress
phonon conductivity. These are currently difficult to compute at the
DFT level due to the high computational demands and empirical force
fields like Tersoff \cite{Tersoff1986} or Brenner \cite{Brenner1990}
potentials are used instead. Parameters for such empirical models are
usually fitted to well understood crystals with simple geometry and
might lack transferability to novel materials with unusual binding
configurations.
In addition, potentials for simulation cells with a larger number of
different elements are scarce. 

In the past years, density-functional theory based tight-binding
(DFTB) \cite{Seifert1986,porezag1995ctb,elstner1998scc} received a lot
of attention, since it provides an intermediate level of theory
between first-principles and empirical methods. In DFTB, the
electronic DFT  Hamiltonian is represented in a reduced atomic orbital
basis and numerically evaluated at a reference density obtained from atomic DFT
calculations. Differences between this reference density and the true
electronic density are accounted for by means of a Taylor-like
expansion. In order to compute total energies, additional pair
potentials (the so-called repulsive potentials) are introduced and
fitted to reproduce the DFT total energy. DFTB has the advantage of a
firm foundation in DFT, while being three orders of magnitude faster
than its parental method.

While there are a couple of investigations that analyze the accuracy
of DFTB vibrational frequencies for finite molecular systems,\cite{Elstner1998,Kruger2005,Gaus2013} phonon
BS have never been systematically studied. The goal of this article is
to provide a benchmark for DFTB and compare to available DFT
results.
Although accurate phonon BS are required in a variety of
fields (e.g., in the interpretation of Raman and infrared spectra or
phase transitions), we
discuss the results with a particular emphasis on possible
applications in thermoelectricity. This is reflected in the selection
of investigated compounds, which are mostly (layered) 2D
materials. Besides the
large availability of phonon BS reference calculations for these now well-studied
systems, this is also motivated by the predicted high figure of merit
of such low-dimensional systems.\cite{Hicks1993,Hicks1993a} To be
specific, we investigate \textit{h}-BN and a selection of carbon based
materials (graphene, graphane and diamond), with the aspiration of
developing a framework for the emerging field of polymer-based
thermoelectricity.\cite{organicthermo} Two allotropes of phosphorene
were also studied to assess the possibilities of going beyond second
row elements. In particular, we interfaced the DFTB+ \cite{dftbp2007} 
implementation of the DFTB method with the phonopy\cite{phonopy} code and apply a general phenomenological approach to compare a descriptor value for thermoelectricity with earlier experimental measurements and theoretical work.

\begin{table*}
\caption{\label{sk} Slater-Koster sets used in this
  study. Hyphened elements in the second row correspond to subsets for
which Slater-Koster files for all element pairs are available. The
fifth row lists the targets: total energies ($E_\text{tot}$), orbital
energies ($\epsilon_i$),
geometries (F) and vibrational frequencies ($\omega$) that were aimed
at during the fitting procedure.}
\begin{ruledtabular}
\begin{tabular}{lcccll}
   SK set  &  Elements & DFTB level & Reference systems &  Targets & Ref.
\\\hline\\
mio-1-1 & O-N-C-H-S-P & 2$^\text{nd}$ order & molecules &
$E_\text{tot}$, F, $\omega$ & [\onlinecite{elstner1998scc}]\\
pbc-0-3 &  Si-F-O-N-C-H, Fe & 2$^\text{nd}$ order &
solids & N/A & N/A \\ 
matsci-0-3 &  Al-O-H, Al-Si-O-H, Cu-Si-Al-Na-O-H, 
& 2$^\text{nd}$ order & molecules &
$\epsilon_i$, F & [\onlinecite{Lukose2010}]\\
& Ti-P-O-N-C-H,  O-N-C-B-H,  Al-O-C-H,  & & &\\
& Si-P-N-O-C-H & & &\\
3ob-3-1 & Br-C-Ca-Cl-F-H-I-K-Mg-N-Na-O-P-S-Zn & 3$^\text{rd}$ order 
& molecules & $E_\text{tot}$, F & [\onlinecite{Gaus2012}]\\
3ob:freq-1-2 & C-C, C-N, C-O & 3$^\text{rd}$ order & molecules &
$E_\text{tot}$, F, $\omega$ & [\onlinecite{Gaus2012}]\\
borg-0-1 & B-N & 2$^\text{nd}$ order & molecules &
$E_\text{tot}$, F, $\omega$ & [\onlinecite{Grundkotter-Stock2012}]
\end{tabular}
\end{ruledtabular}
\end{table*}
\section{Computational Details}
\subsection{DFTB calculations}
DFTB calculations were performed with version 1.2 and 1.3 of the DFTB+
code.\cite{dftbp2007} The geometry optimzations were carried out using a charge
tolerance of $10^{-6}$ in the self-consistent cycle and a maximum force
component of  $10^{-6}$ a.u.. Apart from carbon in the diamond
structure, all other compounds in this study were treated as
monolayers (and hence purely 2D materials) by imposing a unit cell
dimension of 20 \AA\  perpendicular to the layer. We confirmed that
this leads to neglible inter-sheet interactions. 
This choice was made
to allow for a direct comparison to previous computational studies
which often discuss monolayer dispersion relations. Moreover, any
complication due to an insufficient treatment of the Van der Waals
interaction between layers is avoided. All structures were then optimized by constraining
the Bravais lattice to the experimental one and allowing the lattice
constants and basis atoms to relax freely, starting in each case from the
known crystal structure.  Brillouin Zone (BZ) integrations were
carried out using (16 $\times$ 16 $\times$ 1) Monkhorst-Pack (MP) {\bf k}-point
meshes.

In order to perform DFTB calculations,  so-called Slater-Koster
files are required for each element pair in the simulation cell. These
contain tabulated Hamiltonian and overlap matrix elements, as well as
the already mentioned repulsive potentials. The web repository
\url{www.dftb.org} provides a source of currently available
Slater-Koster sets, which have been generated by the DFTB
community. Such sets generally differ in the
actual basis set used to evaluate the DFT Hamiltonian, the highest
order of the Taylor-like expansion around the reference density, and
the reference systems used to create the repulsive potentials.\cite{Elstner2014} In
addition, different groups place more emphasis on the accuracy of
certain properties, like total energies, forces or vibrational modes
in the fitting process.\cite{Gaus2011,Oliveira2015} Table \ref{sk}
lists the Slater-Koster sets used in this study and provides
additional information on their generation. Most sets were generated
for molecular structures and without considering vibrations
explicitly during the parameter generation, the set 3ob:freq-1-2 being a notable exception. The present study
therefore provides a firm test to investigate the transferability of
DFTB as a method, but also the transferability of specific Slater-Koster sets currently used.

\subsection{Phonon-dispersion relations}
DFTB+ has been interfaced to the phonopy code,\cite{phonopy} which
provides a suitable framework to compute phonon BS by the
supercell method (also often referred to as direct method). The new
interface is available in phonopy version 2.1.2. Based on
the primitive unit cell, phonopy creates several supercells with
slightly displaced atoms. In a second step, DFTB+ single-point
calculations are performed on these structures to compute the atomic
forces. These are then collected by phonopy to evaluate the force
constants by numerical differentiation and build the dynamical matrix,
which yields the phonon BS through diagonalization. 

Converged results were obtained by taking the supercell dimension to
be (14 $\times$ 14 $\times$ 1) for graphene and graphane, (16 $\times$
16 $\times$ 1) for \textit{h}-BN and blue phosphorous, (8 $\times$ 8 $\times$ 1)
for black phosphorous and finally (6 $\times$ 6 $\times$ 6) for
diamond. For all supercells, the DFTB single-point calculations were
carried out at the $\Gamma$-point.  

\subsection{Choice of reference}
In order to assess the accuracy of the DFTB phonon-dispersions a
reliable reference needs to be defined. The natural choice would be
experimental data. Since measurements are not always performed at
low-temperature conditions and include anharmonic effects, a direct
comparison to 0 K computations in the harmonic approximation is not
straightforward. For the 2D systems in the present study an additional
complication arises: several compounds have not yet been synthesized as
freestanding monolayers. This influences the band spectra through
interlayer coupling and more importantly through interactions with the
substrate. Hence, we chose first-principles DFT calculations as
reference.         

For molecular vibrations DFT has been extensively benchmarked in the chemistry community.\cite{Johnson1993,Rauhut1995,Finley1995,Hertwig1995} The
hybrid B3LYP and semi-local BLYP exchange-correlation functionals have
emerged as reliable models with average errors of only 20-30 cm$^{-1}$
when appropriate scale factors are introduced.\cite{Scott1996}
Systematic benchmarks for the solid state are much
scarcer. Previous studies\cite{Baroni2001,Hummer2009,He2014} found a
strong dependence of the results on the employed lattice constant. As
an example, the LDA functional provided excellent results when the
same level of theory was used to optimize the structure, but
underestimated phonon modes at the experimental lattice
constant.\cite{Hummer2009}  Given the known tendency of LDA to underestimate
cell volumes, this result can be understood as fortious error
compensation. We finally decided to take the
Perdew-Burke-Ernzerhof\cite{perdew1996gga} gradient-corrected
functional as reference level of theory, mainly because of the large body
of available literature data and the good agreement with experiment (at
the experimental volume) found in Ref.\ [\onlinecite{Hummer2009}]. For graphane
no PBE literature data exists, and we performed or own calculations
using Density-Functional Perturbation Theory as implemented in the
Quantum Espresso suite of programs.\cite{QE-2017}
The corresponding computational parameters are given in the
Supplemental Material.\cite{Suplemental_Thermo}

\begin{table*}
  \caption{\label{latticepara} Lattice parameters of the studied materials at
    different levels of theory in units of \AA.} 
  \begin{ruledtabular}
    \begin{tabular}{lcccccccc}
      \textit{h}-BN & LDA & PBE & HSE & matsci-0-3 & borg-0-1 & Tersoff & empirical & Expt. \\
     \hfill $a$ & 2.494\up{a} & 2.515\up{b} & 2.510\up{c} & 2.550 &
                                                                   2.547  & 2.498\up{d} & 2.505\up{e}  & 2.506\up{f}\\\cline{2-8}
      Diamond &  PBE & matsci-0-3 & pbc-0-3 & mio-1-1 & 3ob-3-1 & 3ob:freq-1-2 & Expt. \\
    \hfill $a$&  3.574\up{g} &  3.583  & 3.562 &  3.558 & 3.600 & 3.615 & 3.567\up{h}            \\
      Graphene \\
    \hfill  $a$& 2.461\up{i} & 2.467 & 2.472 & 2.471 & 2.474 & 2.491 & 2.46\up{j}\\
   Graphane\\
  \hfill $a$&  2.540\up{k} & 2.541 & 2.517 & 2.515 & 2.547 & 2.560 & 2.42\up{l} \\\cline{2-6} 
  Blue phosphorous &  PBE & matsci-0-3 & mio-1-1 & 3ob-3-1 & Expt.\\
    \hfill $a$&  3.326\up{m} & 3.545 & 3.467 & 3.426 &  3.28\up{n} \\
  Black phosphorous \\
  \hfill $a$& 
 N/A & 3.490 & 3.484 &  3.430 & 3.314\up{o}\\
 \hfill $b$& 
 N/A & 4.375 & 4.368 &  4.300 & 4.376\up{o}

    \end{tabular}
  \end{ruledtabular}
\up{a} From Ref.\ [\onlinecite{hBNLDA}];
\up{b} from Ref.\ [\onlinecite{hBNPBE}];
\up{c} from Ref.\ [\onlinecite{CaiHSEhBN}];
\up{d} from Ref.\ [\onlinecite{anees}];
\up{e} from Ref.\ [\onlinecite{Michel2011}];
\up{f} from Ref.\ [\onlinecite{PhysRevB.73.041402}];
\up{g} from Ref.\ [\onlinecite{Hummer2009}];
\up{h} from Ref.\ [\onlinecite{PhysRev.158.805}];
\up{i} from Ref.\ [\onlinecite{graphitepbephonon}];
\up{j} from Ref.\ [\onlinecite{grapheneexplattice}];
\up{k} current work;
\up{l} from Ref.\ [\onlinecite{Elias610}];
\up{m} from Ref.\ [\onlinecite{SUN20162098}];
\up{n} from Ref.\ [\onlinecite{Zhang2016}];
\up{o} from Ref.\ [\onlinecite{Brown:a04860}]
\end{table*} 
\subsection{Harmonic descriptor}
\label{descriptor}
Several measures to quantify the accuracy of DFTB phonon-dispersion relations with respect
to the reference could in principle be imagined. One possibility is to
compare mode frequencies at special points in the BZ. One could also
integrate the difference between DFTB and reference along full bands. Since we are
interested in applications of DFTB in thermoelectricity, we define
instead the following harmonic descriptor for each band $j$:
\begin{equation}
  \label{des}
  D_j = \int_\mathcal{C} \bar{v}_{j}^2({\bf q}) [\hbar
  \omega_{j}({\bf q})]^2 \,dq, 
\end{equation}
where the integral is evaluated along the
high-symmetry lines in the BZ for which the BS is computed. The term $\bar{v}$ is the
group velocity along this line.  This measure is motivated by a comparison with Eq.\ \ref{klat} for the
thermal conductivity. The descriptor $D_j$ incorporates the quantities
entering the thermal conductivity that can be computed already
in the harmonic approximation. Typically, optical phonons contribute
less than acoustic modes to the thermal conductivity although their
phonon frequencies are higher. This is due to a smaller curvature of
the optical bands and hence smaller group velocity, an effect that is
taken into account by the proposed descriptor. 

To determine this descriptor also for the literature data, we first
digitalized the corresponding band structures in the original articles
using WebPlotDigitizer v.\ 4.1.\cite{Rohatgi2018}
 Further, the data
points were interpolated by cubic splines using the routines available
in the SciPy Python library.\cite{Jones2001} The spline representation
also gives direct access to the band derivative. This allowed for a
determination of the group velocities in all cases and gave good
agreement with analytical group velocities computed directly by
phonopy. Eq. \ref{des} was finally evaluated by numerical integration
using the trapezoidal rule with 5000 integration points between any
two special points in the BZ.            

Two measures for the discrepancy between the descriptor value at a
certain level of theory and the reference (in our case DFT with the
PBE functional) were chosen. First, a mean relative error was determined by the deviation of the
descriptor value for each band from the corresponding reference values:
\begin{equation}
\label{criterion_success_abs}
\text{MRE} ={\frac{{\sum_{j=1}^{N}{\big( D_j-D^\text{ref}_j \big)} }}{{\sum_{j=1}^{N}{D^\text{ref}_j}}}},
\end{equation}
where the sum is either over all bands or the subsets of acoustic and
optical bands.   
A mean absolute relative error was applied in parallel, less sensitive to error cancellations:
\begin{equation}
\label{criterion_success_sq}
\text{MARE} =\frac{\sum_{j=1}^{N}{\sqrt{\big( D_j-D^\text{ref}_j
        \big)^2} }}{\sum_{j=1}^{N} D^\text{ref}_j}.
\end{equation}

\section{Results and Discussion}
\subsection{Structural properties}
The structures and unit cells of the studied materials are depicted in
Fig.\ \ref{struct}. Here, \textit{h}-BN, diamond and graphene are
already well known. Monolayer graphane\cite{Sluiter2003}, 
also termed hydrogenated graphene, is a hexagonal
structure with two carbon and two hydrogen atoms in the unit cell. In
the most stable {\em chair} configuration that is studied here, the
hydrogens are alternately adsorbed above and below the graphene
sheet.\cite{Sahin2015} Blue phosphorous is likewise a hexagonal
structure that resembles graphene when viewed upon perpendicular to
the sheet, but is non-planar. This first theoretically predicted
allotrope \cite{Zhu2014b} was later successfully synthesized as single
layer material.\cite{Zhang2016} Single layer black phosphorous,\cite{Liu2014}   also
known as phosphorene, features an anisotropic structure with two
lattice vectors of different length (see Fig.\ \ref{struct}).  

Table \ref{latticepara} summarizes the relevant
lattice parameters obtained at the DFTB level using different
Slater-Koster sets, as well as DFT and experimental literature
data. For \textit{h}-BN the largest body of reference data is
available. Here we also included results from classical MD simulations
using a Tersoff
potential,\cite{Anees2016} and an empirical force constant
model.\cite{Michel2011} Note that due to the limited availability of
DFTB Slater-Koster files for certain elements, not all systems could
be consistently studied with the same sets, matsci-0-3 being an
exception.

For \textit{h}-BN all considered methods agree with each other and
differ from the experimental values by less than 2\%. For
the carbon based materials, pbc-0-3 and mio-1-1 yield nearly identical
structures. The set 3ob:freq-1-2 tends to overestimate lattice
constants. The largest deviation is found for graphane with an error
of 6\%, although the experimental value may be questioned in this
case.  It should be noted that all methods predict an increase of the
lattice constant going from graphene to graphane, contrary to the
experimental results. The phosphor compounds pose larger
problems to DFTB. The matsci-0-3 set overestimates lattice parameters by
8\% in the case of blue phosphorous and 5\% for black
phosphorous. The sets mio-1-1 and 3ob-3-1 likewise overestimate, but
to a smaller degree.      
\begin{figure}[H]
\includegraphics[clip,width=\columnwidth]{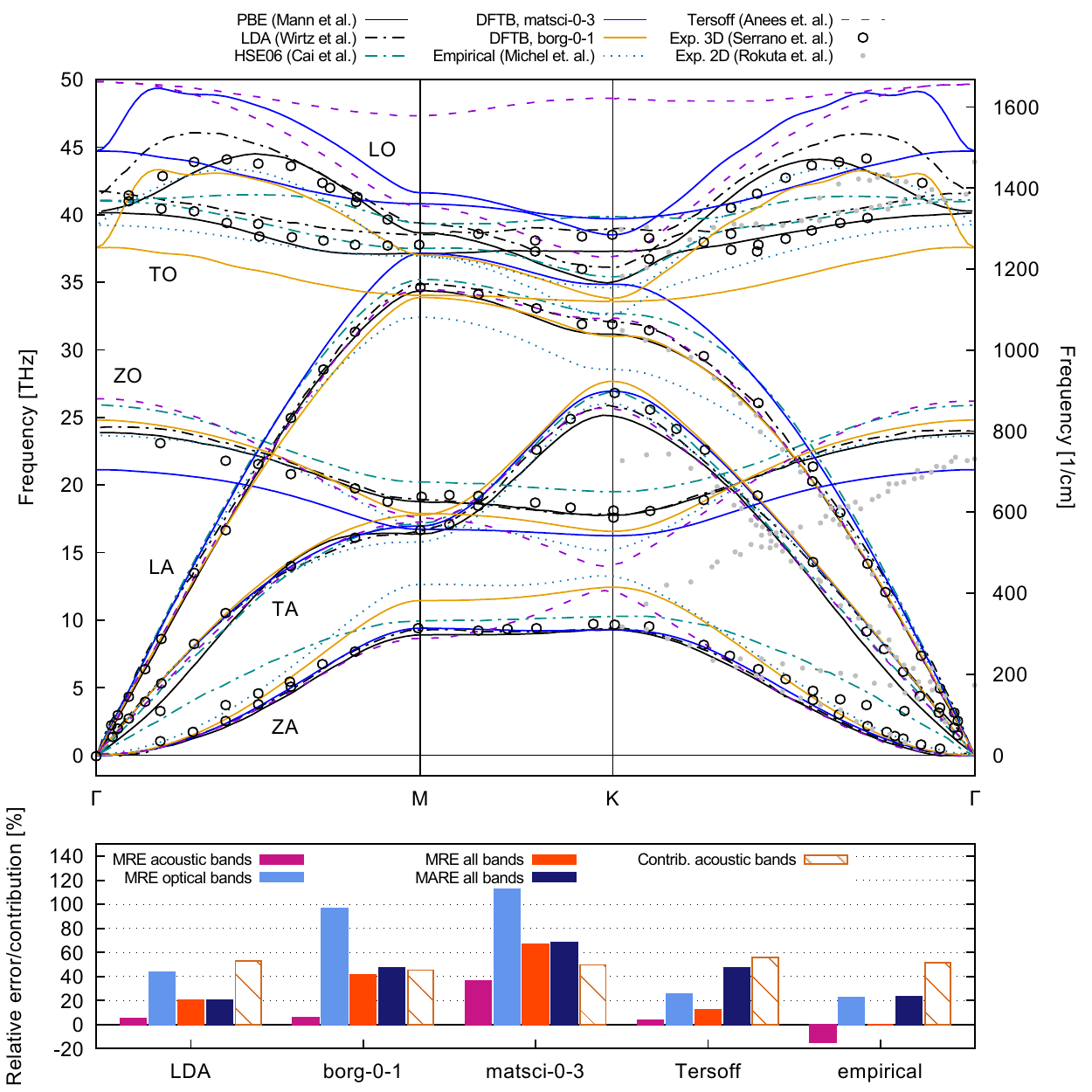}%
\caption{{\bf (Top)} Single layer \textit{h}-BN phonon dispersion.
PBE results from Ref.\ [\onlinecite{hBNPBE}], HSE06 (Ref.\
[\onlinecite{CaiHSEhBN}]), LDA (Ref.\ [\onlinecite{hBNLDA}]), Tersoff
potential (Ref.\ [\onlinecite{anees}]), empirical force field (Ref.\
[\onlinecite{hBNempir}]). The latter and also the HSE06 results are computed at the
experimental geometry, all others use the lattice constant given by
the respective method. Experimental results on
2D-\textit{h}-BN@Ni(111) from Ref.\ [\onlinecite{PRL98_095503}] and
3D-\textit{h}-BN from Ref.\ [\onlinecite{PRL98_095503}]. {\bf (Bottom)}  \textit{h}-BN descriptor errors with
  respect to PBE for the acoustic, optical and the
  total number of bands. The relative contribution of the acoustic bands to
the total descriptor value is likewise given.\label{hBNall}} 
\end{figure}
\subsection{Phonon band structures}

\begin{table*}[!ht]
\caption{\label{table:descriptor} Descriptor values for \textit{h}-BN. 
Mean relative error (Eq. \ref{criterion_success_abs}) with repect to
the PBE reference values for the descriptor of individual bands in \textit{h}-BN.
respectively. "contr." denotes the percentage contribution of the
particular band to the total descriptor value. All values are expressed in \%.
To guide the eye, color markers indicate deviations $<$10 \% (green), to 20\% (yellow), to 50\% (orange), 100\% (red), 200\% (purple) and $>$200\% (black).
}
\begin{ruledtabular}
\begin{tabular}{cccccccccccccccc} & \multicolumn{3}{c}{LDA} &
                                                                   \multicolumn{3}{c}{borg-0-1} & \multicolumn{3}{c}{matsci-0-3} & \multicolumn{3}{c}{Tersoff} & \multicolumn{3}{c}{empirical} \\
band & & error & contr. & & error & contr. & & error & contr. & & error & contr. & & error & contr.  \\
ZA &  \yesq &     11.8 &      0.3  & \blsq &    206.0 &      0.8  &  \yesq &     12.2 &      0.2  &  \pusq &    153.6 &      0.8  & \blsq &    302.0 &      1.4 \\
TA & \grsq &      4.0 &     12.1  &  \orsq &     40.3 &     13.8  &  \orsq &     29.0 &     10.8  & \grsq &      1.9 &     12.7  &  \yesq &     12.5 &     15.7 \\
LA  & \grsq &      4.9 &     40.3  & \grsq &     -6.2 &     30.6  &  \orsq &     39.0 &     38.6  & \grsq &      2.9 &     42.4  &  \orsq &    -25.2 &     34.5 \\
ZO &  \yesq &     19.0 &      2.2  &  \resq &     75.3 &      2.7  &  \orsq &    -45.5 &      0.7  & \blsq &    230.0 &      6.5  &  \resq &     76.2 &      3.9 \\
TO  &  \resq &     52.7 &     21.8  &  \pusq &    121.6 &     26.9  &  \pusq &    156.6 &     26.5  &  \pusq &    130.7 &     35.3  &  \orsq &     28.4 &     22.0 \\
LO  &  \orsq &     38.6 &     23.2  &  \resq &     77.4 &     25.2  &  \resq &     92.0 &     23.2  &  \resq &    -87.5 &      2.2  &  \yesq &     11.7 &     22.4 
\end{tabular}
 \end{ruledtabular}
\end{table*}

{\em \textit{h}-BN} In Fig. \ref{hBNall}, experimentally and computationally
determined phonon band spectra reported in the literature for
\textit{h}-BN are visualized with our results. As usual, acoustic (optical)
longitudinal and transverse modes are denoted as LA (LO) and TA (TO),
respectively. In 2D materials, the flexural modes with atomic
displacements perpendicular to the sheet are further labeled as ZO and
ZA. One can see that the closed, grey symbols representing
experimental results for 
2D-\textit{h}-BN are not reproduced by the different computational
approaches, that are closer to the open symbols, representing
3D-\textit{h}-BN. In practice, 2D-\textit{h}-BN corrugates unpredictably and has to be suspended on a surface to be subjected to analysis.\cite{PRL98_095503}
This means that the experimental 3D-\textit{h}-BN phonon band spectrum
is probably closer to what an idealized experimental 2D-\textit{h}-BN
phonon band spectrum looks like than that of any practical
\textit{h}-BN monolayer. Although 3D-\textit{h}-BN has four atoms in
the unit cell and hence 12 bands are expected in the BS, the weak
inter-layer interaction leads to two sets of essentially degenerate
bands.
As seen in Fig. \ref{hBNall}, larger differences occur
only close to the $\Gamma$-point for the ZA modes and reveal the 3D nature
of the material. 

Discussing first the DFT results, we find that PBE provides an
excellent overall agreement with the experimental data by Serrano et
al.\cite{PRL98_095503} LDA provides the right dispersion throughout
the BZ but overestimates the optical bands. This is opposite to the
typical underestimation given by LDA for other
materials.\cite{Hummer2009,He2014} The HSE results
by Cai\cite{CaiHSEhBN} do not qualitatively reproduce the phonon band
spectrum for the ZA band. The flexural branch should exhibit a quadratic
dispersion as discussed by Carrete et al.\cite{Carrete2016} Reasonable agreement with experiment is found for the DFTB
Slater-Koster sets borg-0-1 and matsci-0-3 for the acoustical
bands. Both show an accurate dispersion for the ZA branch. In fact,
the numerical efficiency of DFTB permits to assess rather larger
supercells and converge also long-range interactions that are
important for finer details of the BS. As another example,
\textit{h}-BN features a maximum of the LO branch away from the
$\Gamma$-point. This overbending is seen in all DFT and DFTB
calculations and due to fifth-neighbour interactions.\cite{Michel2009}
The optical bands of matsci-0-3 are not satisfactory: 
the LO and TA branches are overestimated by around 150 cm$^{-1}$ at
the $\Gamma$-point, and at the same time the ZO branch is underestimated by roughly 100
cm$^{-1}$.

It should be noted that the DFTB approaches feature a second maximum
in the paths $\Gamma \to$ M and K $\to \Gamma$, which is not seen in
the PBE data or the experimental results. We verified that this
feature is due to long-range Coulomb interactions in this weakly
screened material. DFTB zeroth-order simulations, in which there is by
construction no long-range charge-charge interaction, do not show this
behaviour. We believe that the second maximum arises due to an
incomplete treatment of these long-range interactions and could be
overcome by a proper treatment of the nonanalytical part of the
dynamical matrix.\cite{Wang2010} Unfortunately, the required Born
charges and dielectric tensors are not yet implemented in DFTB+, such
that we could not correct the BS at this point. The mentioned artefact
concerns only the optical bands of polar materials in the limit
$\Gamma \to $ 0 and is not expected to influence our general conclusions.

 Turning finally to the empirical approaches, we find an overall
good agreement with experiment for the results of Michel et al.\
\cite{Michel2009}, while the dispersion of the optical bands is
clearly wrong and largely overestimated for the Tersoff potential.   

In order to see how these general trends might influence the lattice
conductivity we now analyze the harmonic descriptor introduced in
Sec.\ \ref{descriptor}. The PBE phonon band structure as determined using
by Mann \textit{et al.}\cite{hBNPBE} was chosen as the
reference to which the other methods were compared. Applying the
descriptor yields the numerical values in Table \ref{table:descriptor}
which are depicted in condensed form also in Fig.\ \ref{hBNall} bottom.

\begin{figure*}[!ht]
 \centering
 \subfloat[Graphene]{%
  \includegraphics[clip,width=0.33\textwidth]{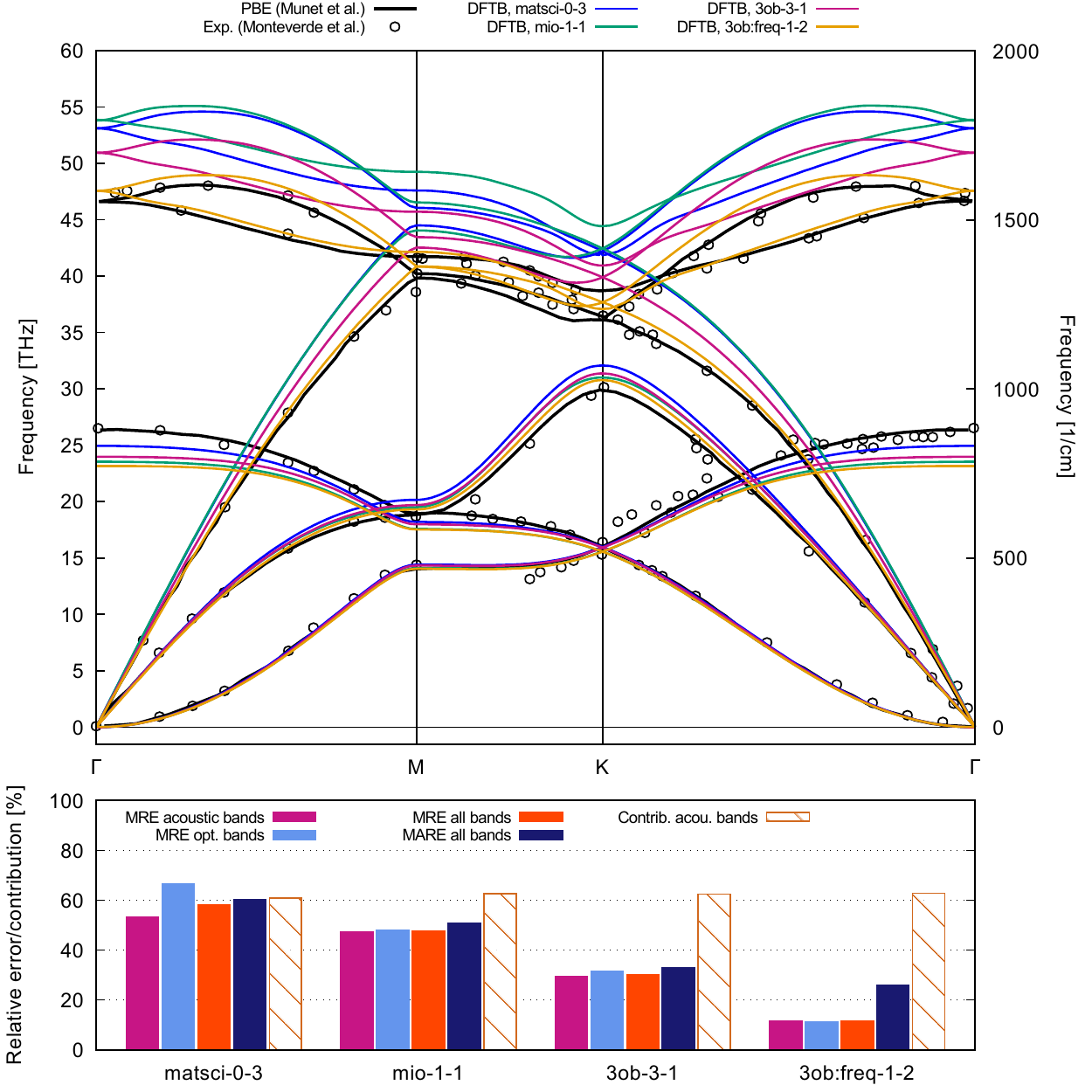}
 }
 \subfloat[Graphane]{%
  \includegraphics[clip,width=0.33\textwidth]{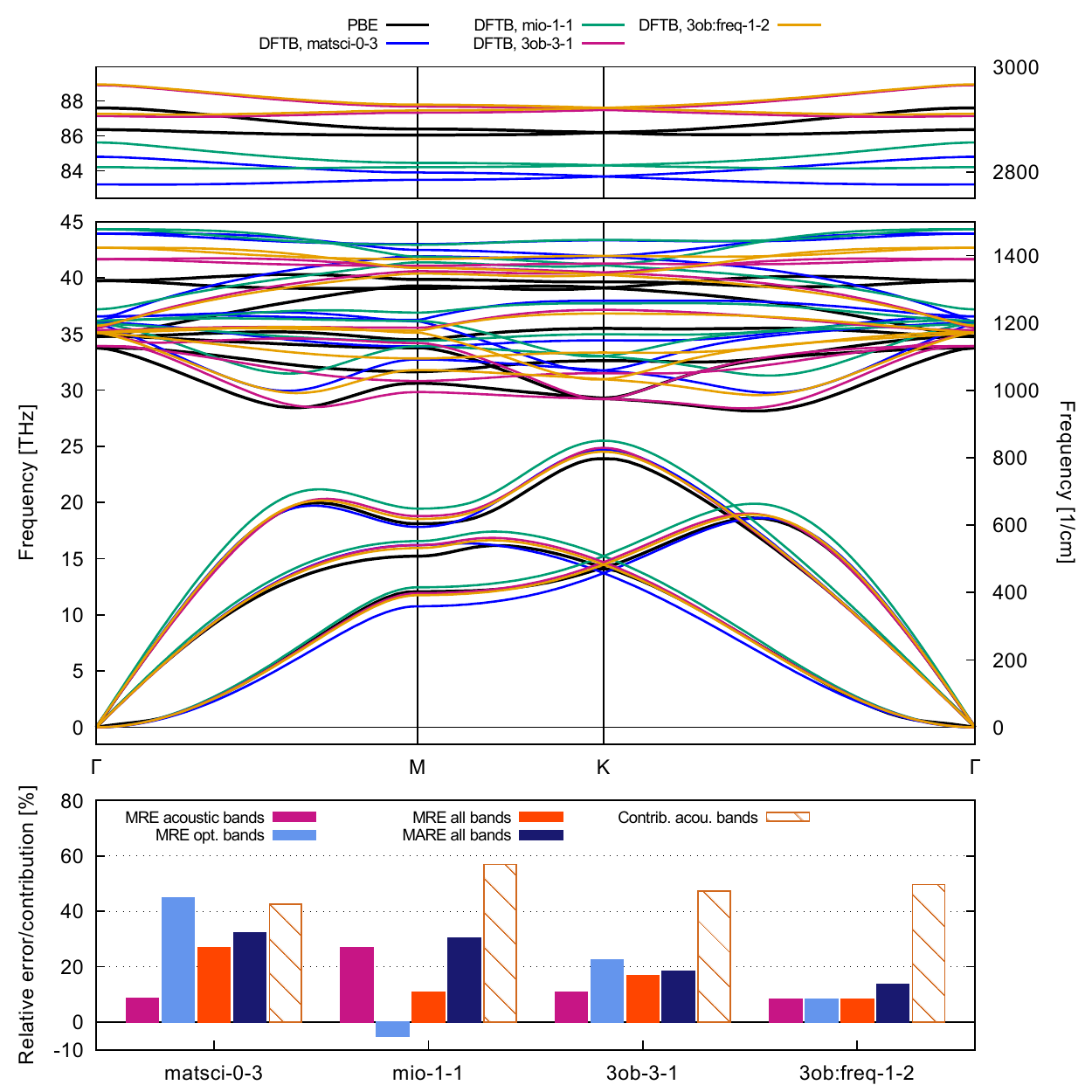}%
 }
 \subfloat[Diamond]{%
  \includegraphics[clip,width=0.33\textwidth]{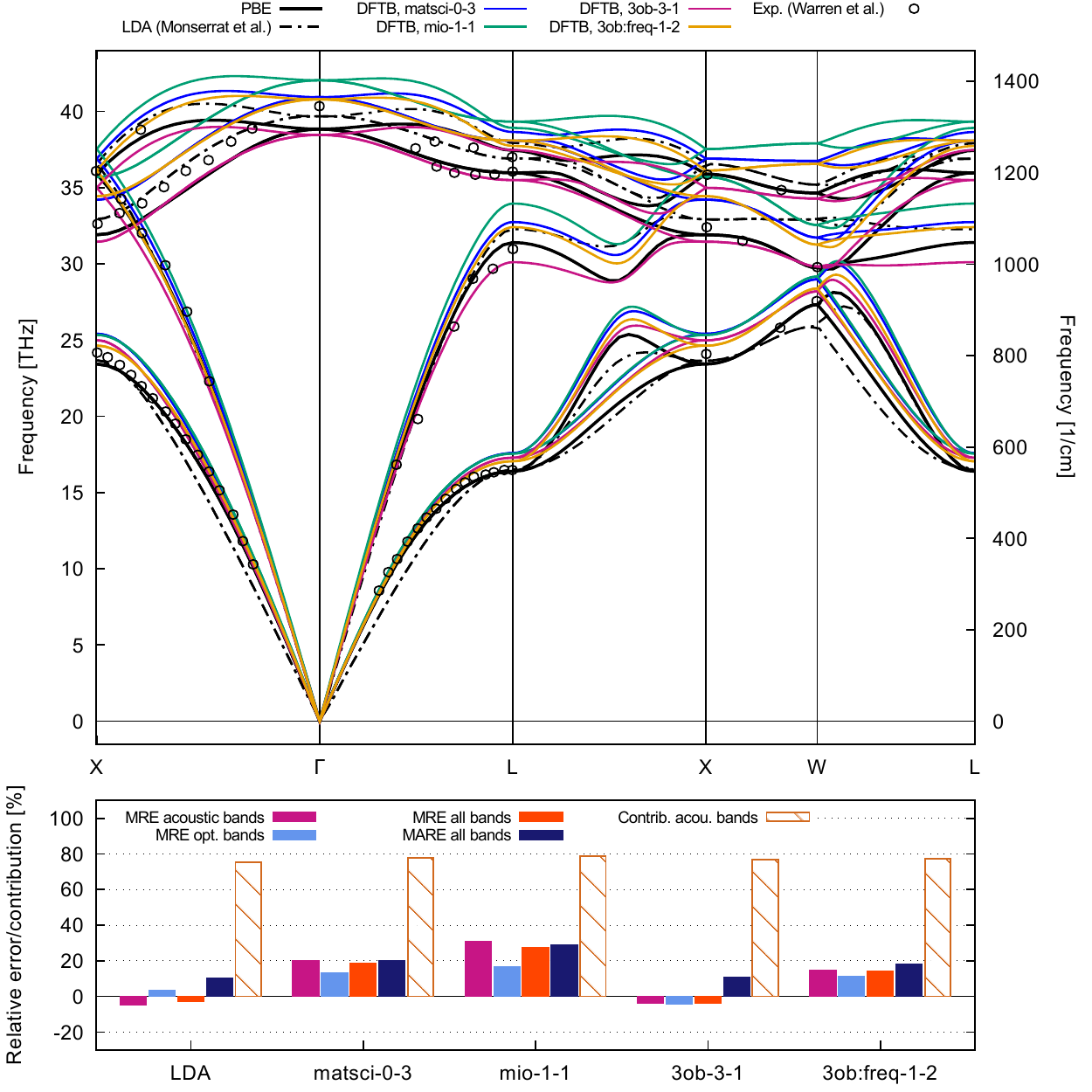}%
 }
   \caption{Phonon dispersion of carbon-based
    compounds. a) Graphene: PBE results from Ref.\
    [\onlinecite{graphitepbephonon}], for the experimental results see Ref.\
    [\onlinecite{Diamondphonons}] and references therein; b) Graphane: PBE results from our
    work, no experimental results available; c) Diamond: PBE results at the experimental lattice constant
    from Ref.\ [\onlinecite{Hummer2009}], experimental results in Ref.\
    [\onlinecite{Diamondphonons}]. If not mentioned otherwise, all
    calculations were performed at the optimized lattice parameter of
    the respective method. Bottom of each graph: descriptor errors and relative
  contribution of acoustic bands.\label{carbonbasedBS}}
\end{figure*}

A crucial feature clear from Table \ref{table:descriptor} is that the
apparent success of a method can be highly sensitive to certain
spectral features, while being insensitive to others. The LA band, for example, alone accounts for $\sim40\%$ of the total value.
Although the difference in frequency of the LA band predicted by PBE and matsci-0-3 at the M-point - where that difference is highest - is less than $10\%$, the sensitivity of the descriptor is such that this translates in a +39\% deviation in descriptor value for that band.
Similarly, the Tersoff potential fails to describe either of the three optical bands in a qualitatively correct fashion; however, the LA and TA bands computed using the Tersoff potential faithfully follow the PBE, and experimental values.
For this reason alone, its descriptor values are better than those of
matsci-0-3, that reproduces the general trends predicted by PBE and
experiment. 
The empirical model performs surprisingly well, although there are
large discrepancies for the descriptor value of the ZA band. Since the
flexural mode contributes only very little ($\sim1\%$) to the total descriptor
value, a rather small total error arises. As expected from the general
earlier discussion, LDA performs well for the acoustic bands but
overestimates the descriptor for the optical bands. The overall MARE
of around 20 \% is still the lowest of all tested methods.

{\em Carbon-based compounds} In Fig.~\ref{carbonbasedBS} the
results for the carbon compounds are shown. A large number of
Slater-Koster sets do include carbon and hence a broader comparison is
possible for this material class. Note that mio-1-1 and pbc-0-3
provided very similar BS and hence only mio-1-1 results are discussed
in the following. We start the discussion with diamond, the only 3D
crystal we considered. PBE follows the experimental phonon dispersion
accurately. Also, all the DFTB models provide correct phonon
dispersions for the acoustic bands, albeit slightly overestimated. The optical bands are also overestimated by around
100 cm$^{-1}$, with 3ob:freq-1-2 giving the smallest error. In total,
the DFTB description of diamond is satisfactory. For graphene the acoustic bands are well reproduced, but the optical
bands are displaced to higher frequencies by slightly over 200 
cm$^{-1}$ for both pbc-0-3 and matsci-0-3. In contrast, 3ob:freq-1-2
yields very accurate BS as found already in an earlier study by Huang
et al.\ \cite{Kuang2015}. Graphane with its additional bands due to
C-H stretch vibrations provides a harder challenge. Compared to the
PBE reference (own calculations), these bands in the 2700 cm$^{-1}$ to
3000 cm$^{-1}$ range are either strongly overestimated (3ob-3-1,
3ob:freq-1-2) or underestimated (matsci-0-3, mio-1-1) by all DFTB models,
while the optical C-C bands are generally overestimated. Again, 3ob:freq-1-2
yields the closest match to the reference.       

This is also seen by considering the harmonic descriptor
(Fig.\ \ref{carbonbasedBS} bottom), which consistently shows the lowest errors for
3ob:freq-1-2 with less than 30 \% in all cases. Matsci-0-3 and mio-1-1
are less
reliable with errors up to 60 \% in the case of graphene. Considering
the sets 3ob-3-1 and 3ob:freq-1-2, we find that 3$^{rd}$ order
DFTB leads generally to an improved description compared to 2$^{nd}$
order DFTB, although the major improvement is seen for 3ob:freq-1-2
which was optimized for frequencies (see Table \ref{sk}). 

Comparing the overall results for graphane and graphene, the larger
errors for the latter material are counter-intuitive. One would think
that due to the structural similarity of both materials the descriptor
errors should also be similar, with graphene having at most lower errors due
the absence of C-H bands. The reason for this unexpected behaviour
is related to the acoustic bands which range up to 1500 cm$^{-1}$ in the
case of graphene, but only up to 800 cm$^{-1}$ for graphane. This
leads to a smaller contribution of the LA and TA bands to the total
graphane descriptor. In addition, the errors of both bands are likewise
smaller for graphane. Another observation is related to   
the set mio-1-1, which shows a negative MRE for the optical bands in graphane. This
can be traced back to the first optical band which contributes the
most to the descriptor of the optical subset. Though mio-1-1 overestimates the
frequencies of this band along the full path, the curvature is much
smaller than the PBE reference, resulting in a smaller descriptor
value. This non-uniform error in optical
vs.\ acoustic bands as well as C-C vs.\ C-H vibrations (see above)
leads to a rather
small MRE for mio-1-1.  Such an error compensation will also be
present in the computation of the final lattice conductivities, but is
clearly system dependent and not desireable. 

The harmonic descriptor also allows to estimate the relative importance of optical
and acoustic bands to the thermal conductivity. While the optical bands contribute only $\approx$
20 \% for diamond, this ratio increases to $\approx$
40-50 \% for graphene and graphane. This highlights the necessity for a
proper description of all modes especially for complex unit cells
with a larger number of optical bands.  

\begin{figure*}[!ht]
 \centering
  \subfloat[Blue phosphorene]{%
  \includegraphics[clip,width=0.5\textwidth]{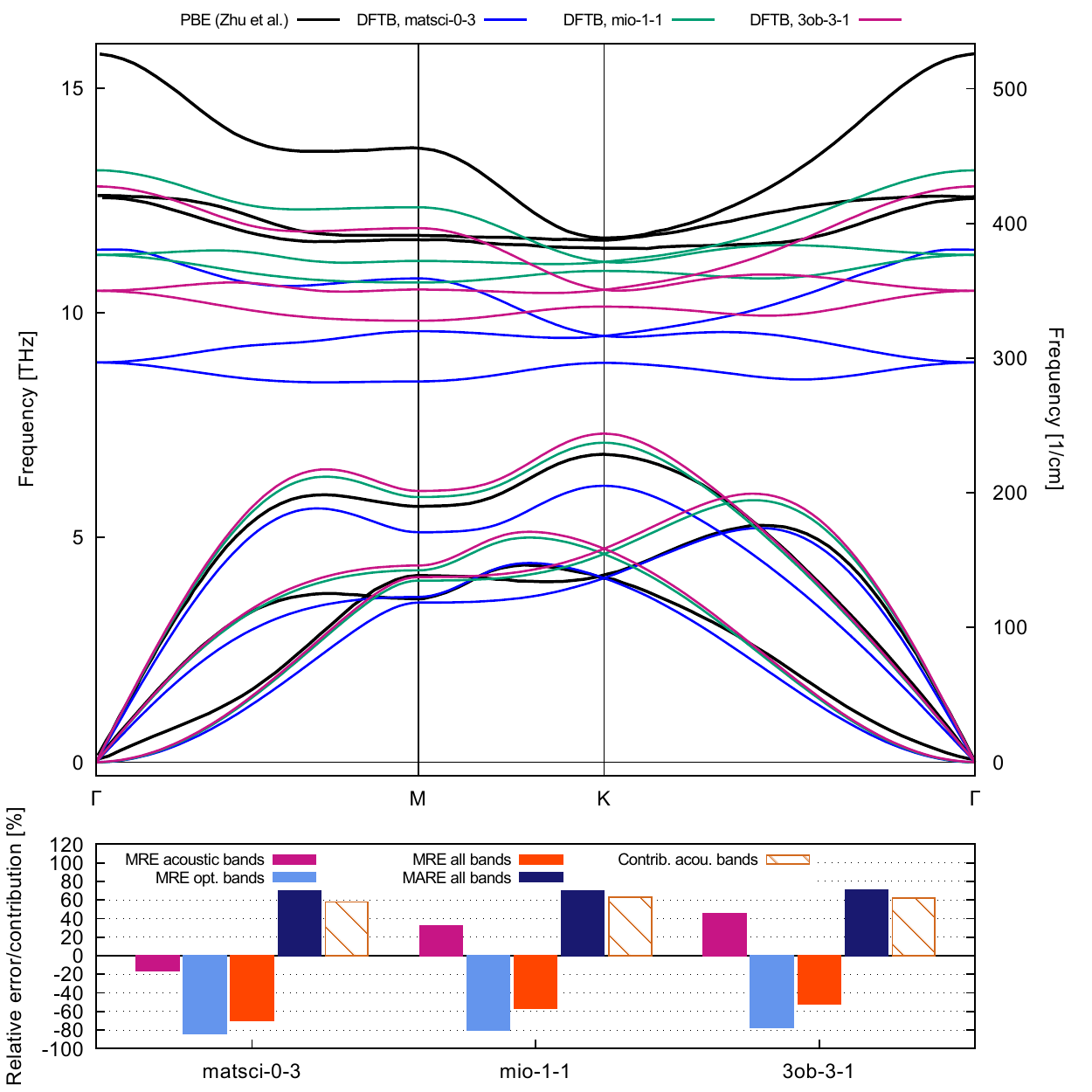}%
  }  
  \subfloat[Black phosphorene]{%
  \includegraphics[clip,width=0.5\textwidth]{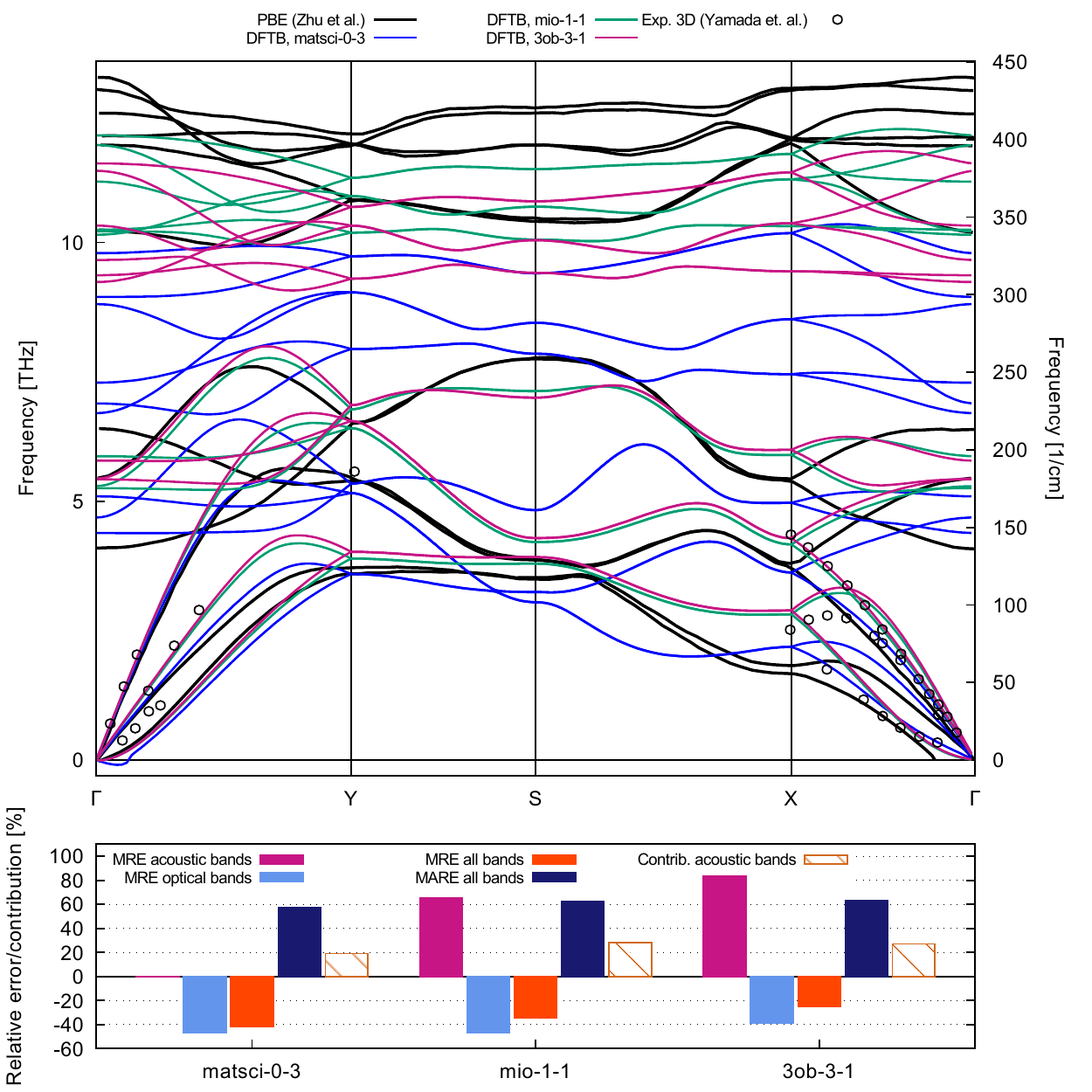}%
  }  
  \caption{Phonon dispersion of 2D phosphorus allotropes.
  For the PBE values, see Ref.\ [\onlinecite{SUN20162098}], optimized lattice parameter.
  For the experimental values of black phosphorus, see Refs.\ [\onlinecite{Yamada1984}]. 
  \label{Pall}}
\end{figure*}

{\em 2D allotropes of phosphorus}
In DFTB one usually employs a minimal basis, taking only those atomic
orbitals into account that are occupied in the respective atom. For
second row elements, like sulfur or phosphorous, this approach leads to
unsatisfactory results because of the hypervalent nature
of bonding in some molecules.\cite{Niehaus2001} As a result, {\em
  d}-orbitals on the second row atom are typically included in the
basis set to improve the results. It is therefore interesting to see
how well DFTB performs for crystalline materials involving
phosphorous.   
      
In Fig.\ \ref{Pall} the results for two allotropes of
2D-phosphorus, black and blue phosphorene, are depicted. Little is known about the experimental phonon band spectrum of black
phosphorus, but the results for blue phosphorene (for which a
comparison to experiment is possible) indicate that PBE is again an accurate
reference. We find for blue phosphorene that all DFTB models strongly
underestimate the optical branch brobdingnagianly. This can be traced back to the
significant overestimation of the lattice constant (Table
\ref{latticepara}). Matsci-0-3 which delivers the largest error in the
crystal structure also underestimates the optical bands by the largest
amount ($\approx$ 30 \%). A similar picture is obtained for black
phosphorene. Here, matsci-0-3 predicts a qualitatively wrong
dispersion with a minimum at S for the optical
band in the 150 cm$^{-1}$ to 300  cm$^{-1}$ range. The Slater-Koster
sets 3ob-3-1 and mio-1-1 perform slightly better in this regard but
also differ strongly from the reference even for the acoustic bands
along the path from S to X.  

It should also be mentioned that regions of negative dispersion are
found for PBE  on
the path X to $\Gamma$ and for matsci-0-3 on the path $\Gamma$ to
Y. The direct approach for computing the phonon band structure is
generally very sensitive to the numerical accuracy of the atomic
forces. For DFTB, we have verified that the results are converged with
respect to k-Point sampling, supercell size and the self-consistent field. In
fact, only the matsci-0-3 SK set shows the mentioned artefact and we
speculate that lower numerical accuracy for the SK tables at long
inter-atomic distance could be the origin.

In Fig.\ \ref{Pall} bottom the numerical results of the
descriptor study clearly show the added difficulty of moving down a
row in the periodic table. The data indicates a strong underestimation
of descriptor values for the optical bands in blue phosphorene. This
is due to the reduced dispersion and lower frequency of these bands in DFTB. As an example, the
highest PBE band has a width of 140 cm$^{-1}$, while mio-1-1
gives a width of only 60 cm$^{-1}$. This also leads to quite
different estimates for the band contribution ratio. While PBE
predicts a very strong contribution of optical bands to the lattice
conductivity with nearly 80 \%, the DFTB values are much lower with
40 \%. Not surprisingly, the total error of the descriptor is
the largest among all systems studied with 70 \% on
average. DFTB with third order corrections (3ob-3-1) is not
significantly better than the other Slater-Koster sets studied. The
special parameterization for frequencies (c.f. 3ob:freq-1-2) is not
available for phosphorous and seems to be an important factor to reach
high accuracy.           

\section{Conclusions}
Summarizing the results of the previous sections, one can say that the
various available Slater-Koster sets provide in most cases accurate crystal
structures and also acceptable acoustic band dispersions. Optical
bands are described less well and can be shifted by several hundred
wavenumbers with respect to the reference, typically to higher
frequencies. Given that the Slater-Koster sets were parameterized for
molecular structures, the overall performance indicates a reasonable degree of
transferabilty. We found also that the accuracy varies strongly for
different Slater-Koster sets. The set 3ob:freq-1-2 clearly outperforms
other available sets, but is clearly limited in the available element
combinations. Judging its quality based on the harmonic descriptor,
3ob:freq-1-2 deviates from PBE by roughly 20 \% (MARE) on average over
the carbon based materials. For comparison, the
LDA results differ also by 20 \% from the reference for
\textit{h}-BN. 
 
Not surprisingly, Slater-Koster sets which were created with molecular
vibrational frequencies as one of the fitting targets perform the
best. This would indicate that new sets covering further elements
should always follow this strategy. Unfortunately, there is evidence that
accurate energetics and vibrations are mutually exclusive
targets. For applications in thermoelectricity this presents no
real problem, since Slater-Koster sets created with a special emphasis
on frequencies do also deliver accurate crystal structures (as we have
shown) which is a prerequisite for proper lattice conductivities. Only
in cases where two phases of the target material are energetically close,
special care is warranted. 

We conclude that the harmonic properties of
2D materials can be successfully computed using the DFTB method at a
fraction of the computational cost of full DFT calculations. This
opens possibilities to perform previously inaccessible phonon
dispersion calculations on thermoelectric polymers and defect engineered
layered materials. Whether
phonon lifetimes from third-order force constants are also sufficiently
accurate is currently under study in our laboratory. 

\begin{acknowledgments}
This project has received funding from the European Union’s Horizon 2020 research and innovation programme under Grant Agreement No 766853.
We would also like to thank the Laboratoire d'Excellence iMUST for
financial support and GENCI for computational resources under
project DARI A0050810637. 
\end{acknowledgments}

\bibliography{Play_new}
\end{document}